\documentclass[11pt,fleqn]{article}

\usepackage{amsmath,amsfonts,amssymb,latexsym,cite}
\usepackage{graphicx}

\def\noi{\noindent}

\newcommand{\Title}[1]{\noi {{\Large\bf #1}}\\[1ex]}

\def\Aunames#1{\noi{\bf #1}}
\def\au#1{${}^{#1}$}
\def\Addresses#1{\medskip\noi \protect
	\begin{description}\itemsep -3pt {\it #1} \end{description}}
\def\adr#1#2{\item[${}^{#1}$]{\it #2}}

\newcommand{\Abstract}[1]{\vskip 2mm \begin{center}
        \parbox{16.4cm}{\small\noi #1} \end{center}\medskip}

\def\email#1#2{\footnotetext[#1]{e-mail: #2}\addtocounter{footnote}{1}}

\def\nq{\hspace*{-1em}}
\def\nqq{\hspace*{-2em}}

\def\qq{\qquad}
\def\cm{\hspace*{1cm}}
\def\inch{\hspace*{1in}}

\def\ten#1{\mbox{$\times 10^{#1}$}}

\def\Jl#1#2{#1 {\bf #2},\ }

\def\ApJ#1 {\Jl{Astroph. J.}{#1}}
\def\CQG#1 {\Jl{Class. Quantum Grav.}{#1}}
\def\DAN#1 {\Jl{Dokl. AN SSSR}{#1}}
\def\GC#1 {\Jl{Grav. Cosmol.}{#1}}
\def\GRG#1 {\Jl{Gen. Rel. Grav.}{#1}}
\def\JETF#1 {\Jl{Zh. Eksp. Teor. Fiz.}{#1}}
\def\JETP#1 {\Jl{Sov. Phys. JETP}{#1}}
\def\JHEP#1 {\Jl{JHEP}{#1}}
\def\JMP#1 {\Jl{J. Math. Phys.}{#1}}
\def\NPB#1 {\Jl{Nucl. Phys. B}{#1}}
\def\NP#1 {\Jl{Nucl. Phys.}{#1}}
\def\PLA#1 {\Jl{Phys. Lett. A}{#1}}
\def\PLB#1 {\Jl{Phys. Lett. B}{#1}}
\def\PRD#1 {\Jl{Phys. Rev. D}{#1}}
\def\PRL#1 {\Jl{Phys. Rev. Lett.}{#1}}

\def\lal{&&\nqq {}}
\def\eq{Eq.\,}

\def\beq{\begin{equation}}
\def\eeq{\end{equation}}
\def\bear{\begin{eqnarray}}
\def\bearr{\begin{eqnarray} \lal}
\def\ear{\end{eqnarray}}
\def\earn{\nonumber \end{eqnarray}}

\def\nnn{\nonumber\\ \lal }

\def\yy{\\[5pt] {}}
\def\yyy{\\[5pt] \lal }

\def\dst{\displaystyle}

\def\fracd#1#2{{\dst\frac{#1}{#2}}}

\def\Half{{\fracd{1}{2}}}

\def\e{{\,\rm e}}
\def\d{\partial}

\def\sign{\mathop{\rm sign}\nolimits}

\def\dim{\mathop{\rm dim}\nolimits}

\def\eps{\varepsilon}
\def\ep{\epsilon}

\def\then{\ \Rightarrow\ }

\def\mn{_{\mu\nu}}
\def\MN{^{\mu\nu}}

\def\cK{{\cal K}}
\def\cV{{\cal V}}

\def\kappa{\varkappa}

\def\wt{\widetilde}
\def\tg{{\wt g}}
\def\tR{{\wt R}}

\def\M{{\mathbb M}}

\def\S{{\mathbb S}}
\def\V{{\mathbb V}}
\def\oR{{\overline R}}

\def\sss{\scriptscriptstyle}
\def\mD{m_{\sss D}}

\def\rf{\eqref}
\def\eqn{\eq\eqref}

\usepackage{color}

\begin{document}
\twocolumn[

\Title{On cosmology in nonlinear multidimensional gravity\yy with multiple factor spaces}

\Aunames{S. V. Bolokhov\au{a,1} and K. A. Bronnikov\au{a,b,c,2}} 

\Addresses{
\adr a {\small Peoples' Friendship University of Russia (RUDN University), 
             ul. Miklukho-Maklaya 6, Moscow 117198, Russia}
\adr b {\small Center for Gravitation and Fundamental Metrology, VNIIMS,
             Ozyornaya ul. 46, Moscow 119361, Russia}
\adr c {\small National Research Nuclear University ``MEPhI''
                    (Moscow Engineering Physics Institute), Moscow, Russia}
	}

\Abstract
  {Within the scope of multidimensional Kaluza--Klein gravity with nonlinear curvature terms and two spherical 
   extra spaces $\S^m$ and $\S^n$, we study the properties of an effective action for the scale factors
   of the extra dimensions. Dimensional reduction leads to an effective 4D multiscalar-tensor theory.
   Based on qualitative estimates of the Casimir energy contribution on a physically reasonable length scale,
   we demonstrate the existence of such sets of initial parameters of the theory in the case $m=n$ that 
   provide a minimum of the effective potential that yield a fine-tuned value of the effective 4D cosmological
   constant. The corresponding size of extra dimensions depends of which conformal frame is interpreted as the 
   observational one: it is about three orders of magnitude larger than the standard Planck length if we adhere
   to the Einstein frame, but it is $n$-dependent in the Jordan frame, and its invisibility requirement 
   restricts the total dimension to values $D = 4+2n \leq 20$. 
  }

]       
\email 1 {boloh@rambler.ru}
\email 2 {kb20@yandex.ru}

\section{Introduction}

 In modern physics, the concept of extra dimensions appears in various theoretical contexts such as 
 Kaluza--Klein models \cite{sb-Kaluza, sb-Bailin, sb-Blagoevic}, brane-world models
 \cite{sb-Randall1, sb-Randall2}, superstring/M-theories \cite{sb-Green}, multidimensional cosmology, etc. 
 Despite the unobservable nature of extra dimensions, their possible existence provides a very elegant 
 theoretical background for a number of key physical problems (geometrization of the interactions, possible
 variations of fundamental constants, etc., see, e.g., \cite{sb-BR-book, sb-KBR1, sb-Mel, sb-ivas}), and it is 
 also an essential and inevitable feature of many theories such as superstring theory \cite{sb-Green}. At present,
 quite a general class of multidimensional theories is under consideration,
 including nonlinear functions of the Ricci scalar ($F(R)$ theories) and high-order curvature invariants (e.g., the
 Gauss---Bonnet term) motivated by various inflationary scenarios and low-energy limits of superstring models.

 One of possible explanations of the unobservable nature of extra dimensions is the concept of spontaneous
 compactification \cite{sb-Bailin, sb-Blagoevic}, according to which the geometry of the entire multidimensional
 manifold (bulk) can be understood as a topological product $\M^4 \times \V^n$ of the observable 4D 
 spacetime  $\M^4$ and a compact extra space $\V^n$ of appropriate dimension and topology with a very 
 small characteristic length scale, in many cases close to Planckian. This kind of explanation is widely used in 
 Kaluza--Klein-like theories. A realistic model of such a type should include (at least in principle) some explanatory
 mechanism for stabilization of the radii of compact extra dimensions to keep them unobservable on a 
 sufficiently large time scale of the Universe evolution. This stability can be global or local, and may be achieved
 on the (quasi)classical or/and quantum level.

 In the simplest cases, the stability conditions of ground state manifolds of the type $\M^4 \times \V^n$ can 
 be related to the existence of minima of an effective potential of scalar fields appearing from the 
 extra-dimensional metric tensor components at dimensional reduction. It is clear that, beyond a classical level,
 there can also be some quantum vacuum contributions to the effective potential such as the Casimir energy 
 of scalar, vector gauge, and spinor fields due to the compact topology of the extra factor space 
 \cite{sb-3, sb-4, sb-5, sb-Candelas}.

 In this paper we continue our study of cosmological models in the framework of nonlinear  
 multidimensional gravity, taking into account possible contributions of quantum vacuum effects. 
 In \cite{sb-16a, sb-16b}, the space-time geometry was chosen in the extended Kaluza-Klein
 form, $\M^4 \times \S^n$, where $\M^4$ is the observed weakly curved 4D space-time, 
 and $\S^n$ is an $n$-dimensional sphere of sufficiently small size to be invisible by modern 
 instruments. Now we extend the study to more complex geometries containing a number 
 (at least two) of internal factor spaces.

 As compared to the Casimir energy on $\M^4 \times \S^n$ manifolds \cite{sb-Candelas, sb-5}, 
 calculations of this energy on $\M^4 \times \S^n \times \S^m$ make a much more complicated problem
 which has been studied in a number of papers \cite{sb-Kikkawa, sb-Gleiser-87}. In the case of an even
 number of extra dimensions, there is an additional difficulty in renormalizing the logarithmically divergent
 terms in the Casimir energy \cite{sb-Birm}.

 The analysis of a possible stability of extra dimensions in multidimensional gravity with high-order curvature
 invariants (up to $R^2$, $R\mn R\MN$, $R_{\mu\nu\alpha\beta}R^{\mu\nu\alpha\beta}$) was performed
 in a pure classical level, without the Casimir contribution, by Wetterich \cite{sb-Wetterich} mostly for 
 $\M^4 \times \S^D$ geometries under a number of assumptions such as neglecting high-order curvature
 contributions to the kinetic terms in $1/D$-approximation and a vanishing effective cosmological constant.
 For geometries $\M^4 \times \S^m \times \S^n$ it was argued that the effective potential may not 
 be bound from below. 

 Our approach to dimensional reduction in theories with nonlinear curvature terms has a number of features
 different from others in some aspects, such as: using a slow-change approximation; a transition from 
 the Jordan conformal frame to the Einstein one; a dynamic treatment of scale factors of extra dimensions
 as scalar fields on $\M^4$ with their own kinetic terms (which need to be checked for positive definiteness);
 a physically reasonable (``semiclassical'') choice of the characteristic length scales of extra dimensions;
 approximate estimation of Casimir contributions at these ranges.

  As in \cite{sb-16a, sb-16b}, we begin with a sufficiently general $D$-dimensional gravitational
  action, then, under suitable assumptions on the space-time geometry, follow a dimensional 
  reduction and a transition to the Einstein conformal frame. After that, we demonstrate the 
  existence of such sets of the initial parameters that provide a minimum of the effective 
  potential at a physically reasonable length scale, and it is also shown that the kinetic term
  of the effective scalar fields is positive-definite, hence these minima really describe stable 
  stationary configurations. 

\section{Basic equations, reduction to 4 dimensions}

 In our description of multidimensional gravity, we follow \cite{sb-BRS-10, sb-BR-book}, where a more 
 detailed derivation can be found. We are considering a $D$-dimensional space-time $\M$ with the structure
\beq
     \M = \M_0 \times \M_1 \times \dots \times \M_N,
\eeq
  where $\dim \M_i = d_i$, and the metric
\beq\label{ds_D}
    ds_D ^2 = g_{ab}(x)dx^a dx^b
		                    + \sum_{i=1}^{N}\e^{2\beta_i(x)} g^{(i)},
\eeq
  where $(x)$ denotes the dependence on the first $d_0$ coordinates $x^a$;
  $g_{ab} = g_{ab}(x)$ is the metric in $\M_0$, and $g^{(i)}$ are
  $x$-independent $d_i$-dimensional metrics of factor spaces $\M_i$,  $i = \overline{1,N}$.

  We are dealing with a theory of gravity with the action
\bearr
     S = \Half m_D^{D-2} \int \sqrt{^D g}\, d^D x                    \label{S_D}
                   \big[ F(R) + c_1 R_{AB} R^{AB} 
\nnn \inch 
		+ c_2 \cK + L_m],
\ear
  where $^D g = | \det(g_{AB})|$, $F(R)$ is an arbitrary function of the scalar curvature 
  $R$ of $\M$; $c_1,\ c_2$ are constants; $R_{AB}$ and $\cK= R_{ABCD}R^{ABCD}$ are the 
  Ricci tensor and Kretschmann scalar of $\M$, respectively; $L_m$ is the matter Lagrangian
  (it can formally include the quantum vacuum contribution).
  Capital Latin indices cover all $D$ coordinates, small Latin ones ($a,b,\ldots$) the coordinates 
  of the factor space $\M_0$, and $a_i,\ b_i, \ldots$ the coordinates of $\M_i$.

  Let us assume that the factor spaces $\M_i$ are $d_i$-dimensional
  compact spaces of constant nonzero curvature $K_i = \pm 1$, i.e., spheres
  ($K_i = 1$) or compact $d_i$-dimensional hyperbolic spaces ($K_i = -1$)
  with a fixed curvature radius $r_0$, normalized to the $D$-dimensional
  analogue $\mD$ of the Planck mass, i.e., $r_0 = 1/\mD$
  (we are using the natural units $c = \hbar =1$). Then we have
\bearr                                                          \label{r0}
    \oR^{a_ib_i}{}_{c_id_i} = K_i\,\mD^2\,\delta^{a_ib_i}{}_{c_id_i},
\nnn
    \oR_{a_i}^{b_i} = K_i\, \mD^2\, (d_i-1) \delta_{a_i}^{b_i}, 
\nnn
         \oR_i = K_i\,\mD^2\, d_i (d_i-1) .
\ear
  The scale factors $r_i(x) \equiv \e^{\beta_i}$ in (\ref{ds_D}) are dimensionless.
  The overbar marks quantities obtained from the factor space metrics
  $g_{ab}$ and $g^{(i)}$ taken separately, $\beta_{,a} \equiv \d_a \beta$,
  and $\delta^{ab}{}_{cd} \equiv \delta_c^a\delta_d^b - \delta_d^a \delta_c^b$ 
  and similarly for other kinds of indices.

  To simplify further calculations, we are using the slow-change approximation \cite{sb-BR-06}:
  we suppose that all derivatives $\d_a$ are small as compared to the extra-dimension scale, 
  so that each $\d_a$ involves a small parameter $\eps$, and we neglect all quantities
  of orders higher than $O(\eps^2)$. This approximation iproves to be valid in almost all 
  thinkable situations. In the descriptions of the modern Universe with small extra dimensions 
  it is valid up to tens of orders of magnitude.

  Then, integrating out all subspaces $\M_i$ in \rf{S_D}
  and subtracting a full divergence, we obtain the $d_0$-dimensional action
\bearr 
     S = \Half {\cV}\, \mD^{d_0-2} \!                           \label{SJ}
                \int \! \sqrt{g_0}\,d^{d_0} x\, \Big \{
                \e^{\sigma} F'(\phi)\oR_0  
\nnn \cm\cm
        + K_J \, -2 \big[V_J(\phi_i) + V_{J(\rm Cas)}\big]\Big\},    
\yyy 	\nq\,				\label{KJ}
     K_J = F'\e^\sigma
            \biggl[-(\d\sigma)^2 + \sum_i \!  d_i(\d\beta_i)^2 -2F'' (\d\phi,\d\sigma)\biggr]\! 
\nnn \cm\cm
     + 4\e^\sigma (c_1 + c_2)\sum_i \! d_i \phi_i (\d\beta_i)^2,
\yyy \nq 								\label{VJ}
     -2V_J (\phi_i) = \! \e^\sigma \biggl[F(\phi)
                   + \sum_i d_i \phi_i^2 \biggl(c_1 + \frac{2c_2}{d_i{-}1}\biggr)\biggr],
\ear
  where $g_0 = |\det(g_{ab})|$, $\cV$ is a product of volumes of $n$ compact
  $d_i$-dimensional spaces $\M_i$ of unit curvature;  
\beq
      \phi_i := K_i m_D^2 (d_i - 1) \e^{-2\beta_i}, \ \ \       \phi := \sum_i d_i\phi_i  ;
\eeq
  $\sum_i$ means $\sum_{i=1}^{N}$; $\sigma := \sum_i d_i \beta_i$;
  $(\d\sigma)^2 \equiv  \sigma_{,a}\sigma^{,a}$,
  $(\d\alpha, \d\beta)= g^{ab}\alpha_{,a}\beta_{,b}$,
  and similarly for other functions; $\Box = g^{ab}\nabla_a \nabla_b$ is the
  $d_0$-dimensional d'Alembert operator; $\oR[g]$ and $\oR_i$ are the Ricci
  scalars corresponding to $g_{ab}$ and $g^{(i)}$, respectively; 
  $F'(\phi) = dF/d\phi$, and $F''(\phi) = d^2F/d\phi^2$; lastly, $V_{J\rm (Cas)}$ is a quantum 
  vacuum (Casimir) contribution to the Jordan-frame potential, to be discussed below.

  The expression (\ref{SJ}) is typical of a (multi)scalar-tensor theory (STT) of gravity in
  a Jordan frame.  For further analysis, it is helpful to pass on to the Einstein frame using
  the conformal mapping
\bearr \label{trans-g}
        g_{ab} \ \mapsto \tg_{ab} = |\e^{\sigma} F'(\phi)|^{2/(d_0-2)} g_{ab},
\ear
 The action (\ref{SJ}) then acquires the form
\bearr                                                      \label{SE}
    S = \Half {\cV}\, \mD^{d_0-2} \!\int\!\! \sqrt{\tg}\, d^{d_0} x
            \Bigl\{ [\sign F'(\phi)]\ \Big[\tR + K_E \Big]
\nnn \inch
       - 2[V_E (\phi_i) +  V_{E (\rm Cas)}] \Bigr\}
\ear
  with the kinetic and potential terms
\bearr                                                             \label{KE}
        K_E =\frac 1 {d_0-2} \biggl(\d\sigma + \frac{F''}{F'}\d\phi\biggr)^2
         + \biggl(\frac{F''}{F'}\biggr)^2 (\d\phi)^2
\nnn \cm
         + \sum_i d_i\biggl[ 1 + \frac{4}{F'} (c_1 + c_2)\phi_i\biggr]
          (\d\beta_i)^2,                                   
\yyy
          - 2V_E (\phi_i) =                                           \label{VE}
            \e^{-2\sigma/(d_0-2)} |F'|^{-d_0/(d_0-2)}
\nnn \qq \times
	\biggl[ F(\phi) + \sum_i d_i \phi_i^2 \biggl(c_1  + \frac{2c_2}{d_i-1}\biggr) \biggr],
\ear
  where the metric $\tg_{ab}$ is used, the indices are raised and lowered with $\tg_{ab}$
  and $\tg^{ab}$, and $V_{E (\rm Cas)}$ is the Casimir contribution to the total Einstein-frame 
  potential obtained from $V_{J(\rm Cas)}$ after the transformation \rf{trans-g}.
  The quantities $\beta_i$ and $\sigma$ are expressed in terms of the $n$
  fields $\phi_i$, whose numbers coincide with the numbers of factor spaces.
   
\bigskip
\section{A search for stable extra dimensions }

\subsection{Equations for $\M_0 \times \S^m \times \S^n$}

  In what follows, we will try to find stable equilibria of the system with the action \rf{VE} that can 
  correspond to the modern state of the expanding Universe with the metric $g_{ab}$, naturally 
  putting $d_0 =4$, so that $\M_0 = \M^4$. We will also restrict ourselves to two extra factor 
  spaces $\M_1$ and $\M_2$. Furthermore, a final interpretation of the results depends on which
  conformal frame is chosen as the physical (observational) one \cite{bm-01, bm-03}, 
  and this in turn depends on the way in which fermions enter into the (so far unknown) underlying
  unification theory of all interactions. From an infinite number of such opportunities, we will consider 
  two most natural ones: the Einstein frame with the action \rf{SE}, and the Jordan frame with the 
  action \rf{SJ}, obtained directly from the D-dimensional theory. 

  Stable points must be found in any case using the action \rf{SE} as minima of the potential 
  \rf{VE} provided that the kinetic term is positive-definite (which is a priory not at all guaranteed).
  Other conditions to be fulfilled by such a minimum are:

\medskip\noi
         A. Since we adhere to classical gravity, the size of the extra dimensions should appreciably exceed
	  the fundamental length scale $r_0 = 1/\mD$, i.e., $r_i/r_0 = \e^{\beta_i} \gg 1$ ($i = 1,2$).

\medskip\noi
         B. The extra dimensions should not be observable by modern instruments, hence,
         $r_i = r_0 \e^{\beta_i} \lesssim 10^{-17}$ cm, which is close to the TeV energy scale. 

\medskip\noi
        C. The 4D cosmological constant $\Lambda_4$ which corresponds to the minimum value of 
        the potential, should be responsible for the observed dark energy density, so that          
\beq                                      \label{sb-fine}
                  0 < \Lambda_4/m_4^2 \sim 10^{-120},
\eeq
       where $m_4 \sim 10^{-5}$ g is the 4D Planck mass.

  Let us assume that our space-time contains two spherical extra factor spaces
  with dimensions $d_1=m$, $d_2=n$, so that $D = 4 + m +n$. Also, for simplicity, we assume that 
\beq
    F(R) = - 2\Lambda_D + R,  
\eeq
  where $\Lambda_D$ is the D-dimensional cosmological constant. 
  Then $ F(\phi) = -2\Lambda_D + \phi$, $F'=1$, $F''=0$, and the curvature nonlinearity of the theory
  is only contained in terms with $c_1$ and $c_2$ in the action \rf{S_D}. Now we can write for the 
  dimensionless quantity $W (x,y) = r_0^2 V_E (x,y)$ and the kinetic term: 
\bearr                       \label{sb-Wmn}
	  W(x,y) = \frac{x^m y^n}{2}\Big[ k_1 x^4 + k_2 y^4   
\nnn \ \ 
           		-  m (m\!-\!1) x^2 - n (n\!-\!1) y^2 + \lambda \Big] + W_{\rm Cas},
\yyy					\label{sb-Kmn}
   	K(x,y) = \frac m2 (\d x)^2 \bigg[ \frac{m+2}{x^2} + 8(m\!-\!1)(C_1\!+\!C_2)\bigg] 
\nnn \cm \ \ 
		+ \frac n2 (\d y)^2 \bigg[ \frac{n+2}{y^2} + 8 (n\!-\!1)(C_1\!+\!C_2)\bigg]
\nnn \qq \inch
		+ \frac{mn (\d x, \d y)}{xy}.
\ear
  where $x \,{=}\, e^{-\beta_1},\ y \,{=}\,e^{-\beta_2}$, $C_1\,{=}\,c_1/r_0^2$,
  $C_2\,{=}\,c_2/r_0^2$,  $\lambda = r_0^2 \Lambda_D$, and
\bearr
	   k_1 = - \frac{m(m-1)}{2}\big[C_1 (m- 1) + 2C_2\big],
\nnn 
	   k_2 = - \frac{n(n-1)}{2}\big[C_1 (n- 1) + 2C_2\big],.
\earn
 
  We will now seek such combinations of the input parameters $m, n, C_1, C_2$ that $W(x,y)$ has a 
  local minimum at some $x=x_0$ and $y=y_0$ much smaller than unity in order to satisfy requirement 
  A. On the other hand, $x_0$ and $y_0$ should not be too small in order to conform to requirement B,
  but the corresponding estimate crucially depends on our assumption on the value of $\mD = 1/r_0$.
  By requirement C, the value of $W$ at such a minimum must be positive but extremely small. 

\begin{figure*} \centering  
       \includegraphics[width=80mm]{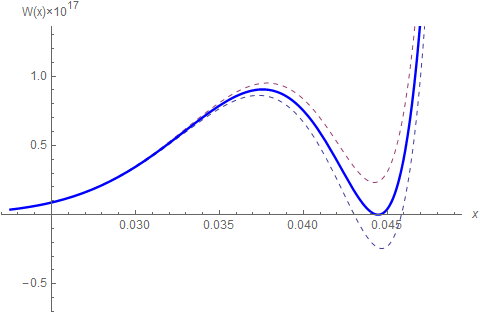}\ \
	\includegraphics[width=80mm]{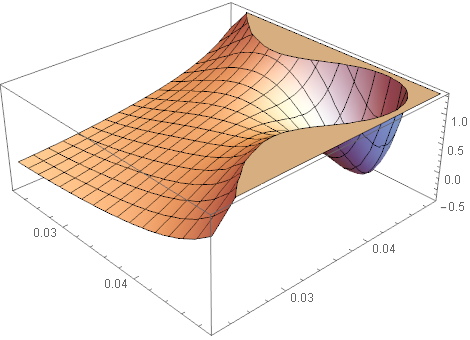}  
\bigskip
      \caption{\protect\small 
	The function $W(x,y)$ close to its minimum for $m=n=5$, $C_1= -127.25$, $C_2 = 128$,
	$\lambda = 0.0197628$. At $x=y$ we have 
	$W = x^{10} (0.0197628 - 20 x^2 + 5060 x^4)$. Left panel: $W(x,x)$ for the parameters 
 	indicated above (solid line), and a bit larger and smaller $\lambda$ (dashed lines). Right panel:
	a 3D plot.}
\end{figure*}
\subsection{The Casimir contribution}
 
 Let us now estimate the Casimir term $W_{\rm Cas}$ in the potential $W(x,y)$. 
 The corresponding calculations are quite complicated and have been performed in a number of
 papers \cite{sb-Kikkawa, sb-Gleiser-87, sb-Birm}
 assuming that the 4D subspace $\M_0$ is flat. Quite evidently, we can use this approximation
 for calculating $W_{\rm Cas}$ if we suppose that $\M_0$ is very weakly curved as compared 
 to the extra factor spaces $\M_1, \M_2$, and it is precisely this approximation that we have 
 been using to obtain all other terms in $W(x,y)$ and $K(x,y)$. Therefore it looks reasonable
 to use the results obtained with flat $\M_0$ in our wider context.

 For estimation purposes we can use the analytic expression for $\M_0 \times \S^3 \times \S^3$
 \cite{sb-Gleiser-87} found for the case of two spheres $\S^3$ of approximately equal radii,
 $r_2/r_1 = 1+ \ep$, where $\ep \lesssim 0.25$, and, in our notations, $r_1 = r_0 e^{\beta_1}$, 
 $r_2 = r_0 e^{\beta_2}$:
\bearr                          \label{sb-VJCas}
        V_{J (\rm Cas)} = - \frac{1}{r_1^4 \cV}\Big[ b \Big(3.639 \ten{-4} - 6.053 \ten{-4} \ep
\nnn \qq
		 + 3.315\ten{-4} (1-2\ep) \ln(r_1/\bar{\mu})\Big)
\nnn \qq
			+ f (3.657\ten{-6}  - 5.414 \ten{-5}\ep) \Big], 
\ear
 where $b$ and $f$ are the numbers of spin-0 and spin-1/2 fields, and $\bar{\mu}$ is an unknown 
 constant emerging in the renormalization procedure. From other calculations, which are mostly
 numerical \cite{sb-Kikkawa, sb-Gleiser-87},
 it follows that $V_{J (\rm Cas)}$ has approximately the same order of magnitude as in \eqn{sb-VJCas},
 that is, $10^{-4}$ or less (due to the factor $1/\cV$) times $r_1^{-4}$ (or, restoring the symmetry 
 between $r_1$ and $r_2$, times $r_1^{-2} r_2^{-2} = r_0^{-4} x^2 y^2$) times the number of 
 field degrees of freedom. The latter may be probably estimated as being of the order of 100. 
 Assuming that the logarithmic term, appearing in the cases of even $m+n$, is not too large (see 
 different arguments on this subject in \cite{sb-Birm}),\footnote
	  {A numerical study shows that if the logarithmic term is large enough to make significant the
	   Casimir contribution to $W$, then a possible minimum of $W$ happens to be with
          $x$ or $y$ close to unity, where our semiclassical approach is no more applicable.}	
 one can write  
\[
   	V_{J (\rm Cas)} \lesssim r_0^{- 4} x^2 y^2,
\]
 and accordingly for the contribution to $W(x,y)$
\beq                         \label{sb-WCas}
		W_{\rm Cas}  \lesssim  x^{2m+2} y^{2n+2}. 
\eeq
  Comparing this expression with \rf{sb-Wmn}, we see that the contribution \rf{sb-WCas}
  contains an extra factor $x^m y^n$, which, provided that other coefficients in \rf{sb-Wmn}
  are of the order of unity, makes the term  $W_{\rm Cas}$ insignificant in a search for a 
  minimum of $W$ at which $x \ll 1$ and $y \ll 1$.  
  
\subsection{Viable minima of $W(x,y)$}

\begin{figure*} \centering  
        \includegraphics[width=80mm]{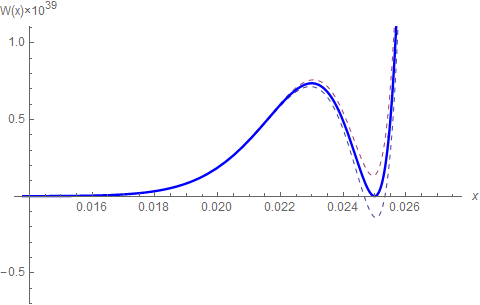}\ \
        \includegraphics[width=80mm]{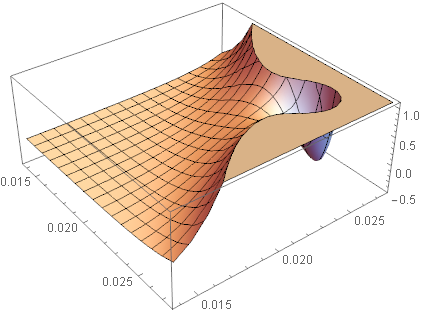}
        \caption{\protect\small 
	The function $W(x,y)$ close to its minimum for $m=n=11$, $C_1= -100$, $C_2 = 100$,
	$\lambda = 11/320$. At $x=y$ we have 
	$W = (11/320)\, x^{22} (1 - 1600 x^2)^2$. Left and right panels: the same as in Fig.\,1.}
\end{figure*}
  In accord with the above-said, we seek a minimum of $W$ ignoring the Casimir contribution. 
  If we additionally assume $m = n$, the expressions \rf{sb-Wmn} and \rf{sb-Kmn} are symmetric 
  with respect to $x$ and $y$, and it makes sense to seek a minimum of $W$ on the line $x=y$,
  which substantially simplifies the process. It turns out that the case $n=3$ is degenerate
  because, instead of two parameters $C_1$ and $C_2$, both $W$ and $K$ depend on the 
  single combination $C_1+C_2$. It turns out that in this case there is no minimum of $W$ combined 
  with a positive-definite $K$, which confirms a previously obtained result \cite{sb-Gleiser-87}.
  
  For other values of $m=n$ it is possible to find a stable minimum of $W$ under the condition 
  $x=y$, for which the kinetic term $K$ can be shown to be positive-definite under the condition
  $c_1+c_2 \geq 0$. It turns out that this minimum occurs at proper ``semiclassical'' values of 
  $x_0 =y_0 \sim 0.01$ under a proper choice of $c_1$ and $c_2$. Moreover, we have 
  obtained analytically such a value of $\lambda$ that $W(x_0) =0$, which, under proper fine
  tuning, makes it possible to satisfy Requirement C:
\beq
		\lambda = - \frac {n(n-1)}{4 C_1 (n-1) + 8 C_2}.
\eeq       
  Evidently, in fact, we need a minimum with $W=0$ in rather a rough approximation: it must be 
  corrected for a corresponding value of $W_{\rm Cas}$ and fine-tuned to obtain a cosmologically 
  relevant $\Lambda_4$, see the next section. 

  Examples of the behavior of $W(x,y)$ leading to viable minima of $W(x,y)$ are shown in Figs.\,1, 2.   
  It has been directly verified that the kinetic term is positive-definite in a neighborhood of the 
  minimum under the condition $C_1+C_2 \geq 0$, which holds in both these examples. 

\bigskip
\section{4D gravity}  
  In multidimensional gravity with constant extra dimensions, the corresponding 4D theory will 
  be evidently the Einstein theory with certain values of the gravitational constant $G_4$ (or the 
  corresponding Planck mass $m_4 = G_4^{-1/2}$) and the effective cosmological constant $\Lambda_4$.
  However, their values expressed in terms of the initial parameters of the theory \rf{S_D} depend on
  which conformal frame is regarded the physical (observational) one. We will consider two most natural 
  opportunities described above, the Einstein frame with the action \rf{SE} and the Jordan frame with
  the action \rf{SJ}. Note that for our present stable extra dimensions we have $m=n$ and $x_0=y_0$,
  therefore $W = W(x_0, y_0) \equiv W(x_0)$.

\subsection{The Einstein frame}

  In the Einstein frame, by \rf{SE}, the 4D Planck mass is $m_4 = \sqrt{\cV(n)} \mD$, and
  $r_0 = \sqrt{\cV}/m_4$. Hence for $x_0 = y_0 \sim 0.01$ the size of extra dimensions is 
  $r(x_0) = r_0/x_0 \sim \sqrt{\cV}\,m_4^{-1}$, close to the Planck length $1/m_4 \approx 
  8\ten{-33}$ cm as long as $\cV(n)$ is not very far from unity.\footnote
	{We have $\sqrt{\cV} \approx 31$ for $n=5$ and $\sqrt{\cV} \approx 16$ for $n=11$.}
  As a result, $r_0 \sim 10^{-31}$ cm, and $r(x_0) \sim 10^{-29}$ cm, manifestly satisfying
  our requirement B that the extra dimensions should be invisible.

  The effective cosmological constant is $\Lambda_4  = W(x_0)/r_0^2 = W(x_0) m_4^2/\cV$,   
  and to conform to observations that require $\Lambda_4/m_4^2 \sim 10^{-120}$, 
  we must have roughly $W(x_0) \sim 10^{-117}$. Therefore, fine tuning is necessary: the
  dimensionless parameter $\lambda$ should be close to the value at which $W(x_0)=0$
  with an accuracy depending on $m = n$. More precisely, according to \rf{sb-Wmn}, we have 
  (for $m=n$ and $x=y$)
\[
                  W (x,y) = \Half x^{2n}(\lambda + {\rm other\ terms}), 
\]
  hence $\lambda$ should be fine-tuned up to $10^{-117} x_0^{-2n}$, where $x_0$ is also 
  different for different $n$. Thus, for $n=5$ (Fig.\,1), the fine tuning of $\lambda$ must be about 
  $10^{-104}$.  For $m=11$ (Fig.\,2), the corresponding accuracy is $\sim 10^{-81}$.  

  The Casimir contribution to $W(x_0)$ is very large as compared to $10^{-120}$: e.g., with the above
  parameter values we have $W_{\rm Cas} \sim 10^{-24}$ for $n=5$ and $\sim 10^{-48}$ for $n=11$ 
  This comparatively large value (though decreasing with growing $m=n$) is compensated 
  by fine-tuned values of other parameters of the theory, above all, $\lambda$.  
  
\subsection{The Jordan frame}

  In the Jordan frame, by \rf{SJ}, the 4D Planck mass is related to $\mD$ by
\bearr                                                                                                \label{r0J}
	\mD^2 = 1/r_0^2 = m_4^2 x_0^{2n}/\cV 
\nnn \cm
\ \then 
       r_0 = \sqrt{\cV} m_4^{-1} x_0^{-n},  
\ear
  and the size of both $\M_1$ and $\M_2$, $r_1 = r_2 = r_0/x_0$. 
  Since $x_0 \ll 1$, $r_0$ is in general a few orders of magnitude larger than the Planck length, 
  which at large enough $n$ may be in tension with the invisibility of extra dimensions.
  And indeed, we obtain:
\bearr  \nq\,
	n=5: \ \  r_0 \approx 1.5\ten{-24}\,{\rm cm}, \ \ r_1\! \approx 3.4\ten{-23}\, {\rm cm};
\nnn  \nq\,
	n=8: \ \  r_0 \approx 1.1\ten{-19}\,{\rm cm}, \ \ r_1\! \approx 3\ten{-18}\, {\rm cm};
\nnn  \nq\,
	n=\! 11{:} \ \  r_0 \approx 5,4\ten{-14}\,{\rm cm}  , \ \ r_1\! \approx 2.14\ten{-12}\,{\rm cm}.
\earn
  Thus $n=5$ and $n=8$ lead to acceptable values of $r_0$ and $r_1$, but at $n>8$ they are too large. 

  The effective cosmological constant in the Jordan frame is obtained if we 
  present the integrand in \rf{SJ} as $\sqrt{g_4}\e^{\sigma} F' [R_4 - 2\Lambda_4  
  + \mbox{kinetic term}]$, which in our case ($F'=1$, $m=n$, $x_0=y_0$) leads to 
  $\Lambda_4  = x_0^{-2n} W(x_0)/r_0^2$. However, expressing $r_0$ in terms of $m_4$, we arrive 
  again at the expression $\Lambda_4 = W(x_0) m_4^2/\cV$. Thus we need the same fine tuning  
  of $\lambda$ as in the Einstein frame and have the same estimate of the Casimir contribution to $W(x)$, 
  despite another value of the fundamental length $r_0 = 1/\mD$. 

\section{Conclusion}

 Considering multidimensional gravity with the action \rf{S_D} in space-times $\M^4 \times \S^m\times \S^n$,
 we have found stable states of the extra dimensions under the assumption $m=n$, and these states 
 are located on the line $x_0=y_0$, which means equal radii of $\S^m$ and $\S^n$. Our attempts to find
 asymmetric stable states (such that $x_0 \ne y_0$) in the case $m=n$, or those corresponding to 
 $m \ne n$, had no success (the corresponding kinetic terms turned out to be not positive-definite), 
 but this is not a strict result, and the existence of such states is not completely excluded.  

 The resulting 4D theory coincides with general relativity when using both the Einstein and Jordan frames 
 to be compared with observation, and in both case we need the same fine tuning of the initial parameters
 of the $D$-dimensional theory in order to have an acceptable value of the cosmological constant 
 $\Lambda_4$. In both cases the fine tuning is slightly weaker than in the ``usual'' cosmological constant 
 problem. However, in these two frames we obtain substantially different estimates of the $D$-dimensional
 Planck scale $\mD$ and the size of extra dimensions $r_1$: while in the Einstein frame $\mD$ almost
 coincides with the conventional Planck mass $m_4$ and $r_1$ is small enough for any $n$, in Jordan's 
 frame the estimated values of $\mD$ and $r_1$ are strongly $n$-dependent, and acceptable results
 are obtained for only $n \leq 8$, that is, $D \leq 20$.     
 
 The dimensional reduction of our models results in classical general relativity well describing the modern state 
 of the Universe and thus far successfully passing all experimental tests. In cosmological applications, our 
 models actually lead to the Einstein equations with nonzero $\Lambda_4$, used in the ``concordance'' 
 $\Lambda$CDM model, but a more advanced problem of describing the whole history of the Universe
 is not addressed. It may seem to be a step back as compared to many studies that try to give such a 
 description, in particular, those using the same multidimensional action \rf{S_D}. 

 Thus, in \cite{sb-BRS-10} it was shown that in a model with two extra factor spaces, one can choose the
 initial parameters of the theory in such a way that the resulting model describes an early inflationary stage 
 (with rolling down along a comparatively steep slope of an effective potential of two scalars $\phi_1$ and 
 $\phi_2$) and that of modern accelerated expansion described by an extremely slow descent along a 
 shallow valley of the same potential, with slow increase of the extra dimensions within observational limits.
 However, unlike ours, the model of \cite{sb-BRS-10} did not take into account quantum vacuum effects,
 which are comparatively easily included in a description of the modern stage (maybe actually beginning at
 the end of the early inflation) but are strong and make a serious problem at the earliest stage, where the 
 extra dimensions should be highly nonstationary. This goes beyond the scope of our present study, but a
 qualitative picture may look as follows. The Universe emerges from a large fluctuation of space-time foam
 in a state corresponding to a point somewhere above the minimum of $W(x,y)$ (if some analog of such a 
 potential can be built in a more advanced model). Then comparatively rapidly the fields roll down to the minimum,
 and, before they settle down there, decaying oscillations around the minimum give rise to creation of matter.
 Such a scenario actually conforms to chaotic inflation. This construction may be a subject of future work.
 
\subsection*{Acknowledgments}

  We thank Milena Skvortsova for helpful discussions.  
  This publication has been prepared with the support the RUDN University Program 5-100
  and by RFBR grant 16-02-00602. The work of KB was also partly performed within the
  framework of the Center FRPP supported by MEPhI Academic Excellence Project 
  (contract No. 02.a03. 21.0005, 27.08.2013).


\end{document}